\documentclass[journal, twocolumn, 10pt, twoside]{IEEEtranTCOM}

\normalsize
\IEEEoverridecommandlockouts
\usepackage{cite}
\usepackage{amsmath,amssymb,amsfonts}
\usepackage{algorithm}
\usepackage[noend]{algpseudocode}
\usepackage{graphicx}
\usepackage{textcomp}
\usepackage{xcolor}
\usepackage{multirow}
\usepackage{boldline}
\usepackage{amsmath}
\usepackage{amssymb}
\usepackage{bm}
\usepackage{hyperref}

\newcommand{\A}{\mathcal{A}}
\newcommand{\C}{\mathcal{C}}
\newcommand{\bmy}{\bm{y}}
\newcommand{\bms}{\bm{s}}
\newcommand{\bmh}{\bm{h}}
\newcommand{\Brace}[1]{\left\{#1\right\}}

\newcommand{\SLVD}{\text{SLVD}}

\newcommand{\SSV}{\text{SSV}}
\newcommand{\trace}{\text{trace}}
\newcommand{\List}{\text{list}}
\newcommand{\WAVA}{\text{WAVA}}
\newcommand{\E}{\mathsf{E}}

\DeclareMathOperator*{\IEE}{IEE}

\usepackage{array}
\newcolumntype{P}[1]{>{\centering\arraybackslash}p{#1}}

\newtheorem{theorem}{Theorem}
    
\begin{document}

\title{
CRC-Aided High-Rate Convolutional Codes With Short Blocklengths for List Decoding\\
}


\author{Wenhui~Sui,~\IEEEmembership{Student Member,~IEEE,}    
Brendan~Towell,~\IEEEmembership{Student Member,~IEEE,}
Ava~Asmani,~\IEEEmembership{Student Member,~IEEE,}
Hengjie~Yang,~\IEEEmembership{Member,~IEEE,}
Holden~Grissett,~\IEEEmembership{Student Member,~IEEE,}
and~Richard~D.~Wesel,~\IEEEmembership{Fellow,~IEEE}
\thanks{ 

This research was supported by the National Science Foundation (NSF) under Grant CCF-2008918. An earlier version of this paper was presented in part at the 2022 IEEE International Conference on Communications (ICC) \cite{Sui2022}.

Wenhui Sui, Brendan Towell, Ava Asmani, Holden Grissett, and Richard D. Wesel are with the Department of Electrical and Computer Engineering, University of California at Los Angeles, Los Angeles, CA, 90095 USA (e-mail: wenhui.sui@ucla.edu; brendan.towell@ucla.edu; ava24@ucla.edu;holdengs@ucla.edu; wesel@ucla.edu).

Hengjie~Yang is with Qualcomm Technologies, Inc., San Diego, CA 92121, USA (e-mail: hengjie.yang@ucla.edu).
}.}

\markright{IEEE Transactions on Communications}

\maketitle

\begin{abstract}
Recently, rate-$1/n$ zero-terminated (ZT) and tail-biting (TB) convolutional codes (CCs) with cyclic redundancy check (CRC)-aided list decoding have been shown to closely approach the random-coding union (RCU) bound for short blocklengths. This paper designs CRC polynomials for rate-$(n-1)/n$ ZT and TB CCs with short blocklengths. This paper considers both standard rate-$(n-1)/n$ CC polynomials and rate-$(n-1)/n$ designs resulting from puncturing a rate-$1/2$ code. The CRC polynomials are chosen to maximize the minimum distance  $d_{\min}$ and minimize the number of nearest neighbors  $A_{d_{\min}}$. For the standard rate-$(n-1)/n$ codes, utilization of the dual trellis proposed by Yamada \emph{et al.} lowers the complexity of CRC-aided serial list Viterbi decoding (SLVD). CRC-aided SLVD of the TBCCs closely approaches the RCU bound at a blocklength of $128$. This paper compares the FER performance (gap to the RCU bound) and complexity of the CRC-aided standard and punctured ZTCCs and TBCCs.
This paper also explores the complexity-performance trade-off for three TBCC decoders:  a single-trellis approach, a  multi-trellis approach, and a modified single-trellis approach with pre-processing using the wrap around Viterbi algorithm.  
\end{abstract}

\section{Introduction}

The structure of concatenating a convolutional code (CC) with a cyclic redundancy check (CRC) code has been a popular paradigm since 1994 when it was proposed in the context of hybrid automatic repeat request (ARQ) \cite{Rice1994}. It was subsequently adopted in the cellular communication standards of both 3G \cite{3GPP2006} and 4G LTE \cite{3GPP2018}. In general, the CRC code serves as an outer error-detecting code that verifies if a codeword has been correctly received, whereas the CC serves as an inner error-correcting code to combat channel errors.

Recently, there has been a renewed interest in designing powerful short blocklength codes. This renewed interest is mainly driven by the development of finite blocklength information theory by Polyanskiy \emph{et al.}, \cite{Polyanskiy2010} and the stringent requirement of ultra-reliable low-latency communication (URLLC) for mission-critical IoT (Internet of Things) services \cite{Ji2018}. In \cite{Polyanskiy2010}, Polyanskiy \emph{et al.} developed a new achievability bound known as the random-coding union (RCU) bound and a new converse bound, known as the meta-converse (MC) bound. Together, these two bounds characterize the error probability range for the best short blocklength code of length $N$ with $M$ codewords. The URLLC for mission-critical IoT requires that the time-to-transmit latency is within 500 $\mu$s while maintaining a block error rate no greater than $10^{-5}$.

Several short blocklength code designs have been proposed in the literature. Important examples include the tail-biting (TB) convolutional codes decoded using the wrap around Viterbi algorithm (WAVA) \cite{Gaudio2017}, extended BCH codes under ordered statistics decoding \cite{Coskun2019, Yue2021}, non-binary low-density parity-check (LDPC) codes \cite{Dolecek2014}, non-binary turbo codes \cite{Liva2013}, and polar codes under CRC-aided successive-cancellation list decoding \cite{Tal2015}. Recent advances also include the polarization adjusted convolutional codes proposed by Ar\i kan\cite{Arikan2019}. As a comprehensive overview, Co{\c s}kun \emph{et al.} \cite{Coskun2019} surveyed most of the contemporary short blocklength code designs in the recent decade. We refer the reader to \cite{Coskun2019} for additional information.

In \cite{Yang2022}, Yang \emph{et al.} proposed CRC-aided CCs as a powerful short blocklength code for binary-input (BI) additive white Gaussian noise (AWGN) channels. The convolutional encoder of interest has rate-$1/n$ and is either zero-terminated (ZT) or TB. 
Yang \emph{et al.} seek to design a \emph{distance-spectrum optimal} (DSO) CRC polynomial that minimizes frame error rate (FER) at a specified signal-to-noise ratio.   For high SNR this is often equivalent to selecting the CRC polynomial that maximizes the minimum distance and minimizes the number of nearest neighbors for the resulting concatenated code.  This design follows the approach of Lou {\em et. al.} \cite{Lou2015} for designing CRC polynomials matched to the CC for improved error detection.

The presence of a CRC facilitates the use of the serial list Viterbi decoding (SLVD), an efficient algorithm originally proposed by Seshadri and Sundberg \cite{Seshadri1994}. Yang \emph{et al.} \cite{Yang2022} showed that the expected list rank of SLVD of the CRC-aided CC is small at high SNR,
thus achieving a low average decoding complexity at operating points of common interest. Yang \emph{et al.} \cite{Yang2022} demonstrated that these concatenated codes can
approach the RCU bound. In \cite{Schiavone2021,Schiavone2023}, Schiavone extended this line of work by looking at the parallel list Viterbi decoding and applying this perspective to legacy coding. In our precursor conference paper \cite{Sui2022}, this framework is extended to rate-$(n-1)/n$ CCs and the resulting concatenated code is able to approach the RCU bound with a low decoding complexity as well.

\subsection{Contributions}
This paper presents designs of CRC-aided ZT and TB CCs for rate-$(n-1)/n$ CCs at short blocklengths for the BI-AWGN channel. Each CRC polynomial is selected to maximize the minimum distance and minimizes the number of nearest neighbors for the resulting concatenated code. If SLVD is performed with a sufficiently large maximum list size such that finding a CRC-passing codeword is guaranteed, then SLVD is an implementation of maximum-likelihood (ML) decoding for these concatenated codes.

We consider standard, systematic, rate-$(n-1)/n$ feedback convolutional encoders \cite{LinCostello2004}, as well as encoders obtained by puncturing the output of a rate-$1/2$ convolutional encoder, as proposed by  Cain {\em et al.} \cite{Cain1979}. The resulting concatenated codes are called a standard or punctured CRC-ZTCC or CRC-TBCC, respectively.

For the standard CCs, SLVD on the primal trellis requires high decoding complexity due to the $2^{n-1}$ outgoing branches at each node. SLVD implementation becomes exponentially more complicated when there are more than two outgoing branches per state. In order to simplify SLVD implementation and reduce complexity, we utilize the \emph{dual trellis} pioneered by Yamada \emph{et al}.\cite{Yamada1983}. The dual trellis expands the length of the primal trellis by a factor of $n$, while reducing the number of outgoing branches at each node from $2^{n-1}$ to at most two. 

For the punctured codes, Cain {\em et al.} \cite{Cain1979} were the first to propose obtaining high-rate CCs through puncturing a rate-$1/2$ code to allow the use of a simple primal trellis for decoding. In \cite{HACCOUN1989}, Haccoun and Bégin presented optimal encoder generators of the original rate-$1/2$ codes and the corresponding puncturing matrices for the rate-$(n-1)/n$ punctured codes. A list-decoding sieve \cite{WeselISTC2023} is used to identify the CRC polynomials that maximize the minimum distance and minimize the number of nearest neighbors for the Haccoun and B\'egin punctured codes.

For ZTCCs, a work related to this line of research is that of Karimzadeh and Vu \cite{Karimzadeh2020}. They considered designing the optimal CRC polynomial for multi-input ZTCCs. In their framework, the information sequence is first divided into $(n-1)$ streams, one for each input rail, and they aim at designing an optimal CRC polynomial for each rail. In contrast, this paper encodes the information sequence with a single CRC polynomial and is then divided into $(n-1)$ streams for the standard rate-$(n-1)/n$ encoder.  Simulation results indicate the new approach improves FER performance at high SNRs.

For TBCCs, simulations show that for both the standard and punctured rate-$(n-1)/n$ TBCCs with blocklength $N=128$, the FER performance of our CRC-aided TBCCs approaches  the RCU bound as the degree $m$ of the CRC polynomial increases.

This paper considers three architectures to enforce the TB condition for CRC-TBCCs. One approach performs list decoding on a single trellis that allows all initial states.  As the list decoder identifies possible trellis paths, non-tail-biting paths are rejected.  At low SNR, SLVD on the single trellis requires a large list size to identify the ML TB codeword, with a majority of trellis paths not satisfying the TB condition. 

An alternative is a multi-trellis approach that initializes 
multiple copies of the dual trellis, one for each possible starting and ending state pair. The multi-trellis approach requires a much smaller list size because only TB paths are added to the list. 
This approach avoids the decoding time complexity of potentially exploring a large list at the cost of the computational and space complexity of creating and storing multiple trellises. It can provide a benefit over the single-trellis approach in low-SNR environments.

A third approach to TB decoding that can be applied in the context of the single-trellis list decoder is the wrap-around Viterbi algorithm (WAVA).  Introduced in \cite{Shao2003}, WAVA is a near-ML decoding algorithm for TBCCs. To achieve the balance of decoding time and space efficiency, this paper combines the wrap-around behavior of WAVA with SLVD for TBCCs. The decoding process is completed in two steps: the WAVA step with at most $2$ trellis iterations, and the list decoding step with a sufficiently large list size such that there are no negative acknowledgment signals. Simulation results demonstrate that this decoding method reduces the average list size as compared with the single-trellis decoder without WAVA, but this reduction comes at a cost of degraded FER performance. 

This paper culminates by providing plots that show the trade-off between decoder complexity and FER performance computed as the gap from the RCU bound at FER $10^{-4}$ and FER vs. SNR results that show both punctured and standard TBCCs performing within $0.1$ dB of the RCU bound with a degree-$10$ CRC polynomial designed to maximize the minimum distance and minimize the number of nearest neighbors.

\subsection{Organization}
The remainder of this paper is organized as follows. Sec.~\ref{sec:SystematicAndDUalTrellis} reviews systematic encoding with feedback for $(n, n-1, v)$ convolutional codes and describes the dual trellis construction. Sec.~\ref{sec: SLVD} introduces various serial list decoders for rate-$(n-1)/n$ CCs. Sec.~\ref{sec: CRC design} presents optimal CRC polynomial designs for standard and punctured high-rate codes using a trellis event enumeration method and a sieve method. Sec.~\ref{sec: complexity analysis} analyzes the trade-off between complexity and decoding performance of these designs. Sec.~\ref{sec: results} presents simulation results comparing FER performance of our designs to the RCU bound and to the Karimzadeh and Vu approach. Finally, Sec. \ref{sec: conclusion} concludes the paper.


\section {The Dual Trellis}
\label{sec:SystematicAndDUalTrellis}
This section describes systematic encoding for $(n, n-1, v)$ convolutional codes and introduces the dual trellis proposed by Yamada \emph{et al.} \cite{Yamada1983} for high-rate CCs generated with an $(n, n-1, v)$ convolutional encoder, where $v$ is the number of memory elements in the encoder. This section also discusses the tree-trellis algorithm for list decoding and its benefits. 

Let $K$ and $N$ denote the information length and  blocklength in bits. Let $R = K/N$ denote the rate of the CRC-aided CC. A degree-$m$ CRC polynomial is of the form $p(x) = 1 + p_1x + \cdots + p_{m-1}x^{m-1} + x^m$, where $p_i\in\{0,1\}$, $i=1, 2, \dots, m-1$. For brevity, a CRC polynomial is represented in hexadecimal where its binary coefficients are written from the highest to lowest order. For instance, 0xD represents $x^3 + x^2 + 1$. The codewords are BPSK modulated. The SNR is defined as $\gamma_s \triangleq 10\log_{10}(A^2)$ (dB), where $A$ represents the BPSK amplitude and the noise is distributed as a standard normal. This SNR can also be written as $2 \frac{E_c}{N_o}=2 R\frac{E_b}{N_o}$, where $E_c$ is the channel symbol energy, $E_b$ is the bit energy, and $\frac{N_o}{2}$ is the two-sided power spectral density of noise.

\subsection{Systematic Encoding}
We briefly follow \cite[Chapter 11, p. 482]{LinCostello2004} in describing a systematic $(n, n-1, v)$ convolutional encoder with reverse input and output labelling\footnote{The reverse input (resp. output) labeling means that the labels from top to bottom of the input (resp. output) streams are arranged in decreasing order.}. A systematic $(n, n-1, v)$ convolutional encoder can be represented by its parity check matrix
\begin{align}
    H(D) = [h^{(n-1)}(D), h^{(n-2)}(D), \dots, h^{(0)}(D)],
\end{align}
where each $h^{(i)}(D)$ is a polynomial of degree up to $v$ in delay element $D$ associated with the $i$-th code stream, i.e., 
\begin{align}
    h^{(i)}(D) = h_{v}^{(i)}D^v + h_{v-1}^{(i)}D^{v-1} + \cdots + h_0^{(i)},
\end{align}
where $h_j^{(i)}\in\{0, 1\}$. For convenience, we represent each $h^{(i)}(D)$ in octal form. For instance, $H(D)=[D^3+D^2+D+1, D^3+D^2+1, D^3+D+1]$ can be concisely written as $H=(17, 15, 13)$. Let $\bm{h}^{(i)} \triangleq [h_v^{(i)}, h_{v-1}^{(i)}, \dots, h_0^{(i)}]$, $i=0, 1, \dots, n-1$. Since the $0$th code stream corresponds to the parity check bit and the $i$th code stream corresponds to the $i$th input bit for $1\le i\le n-1$, the systematic encoding matrix $G(D)$ associated with $H(D)$ is thus given by
\begin{align}
    G(D) = \begin{bmatrix}
    \frac{h^{(1)}(D)}{h^{(0)}(D)} & 1 & 0 & \cdots & 0\\
    \frac{h^{(2)}(D)}{h^{(0)}(D)} & 0 & 1 & \cdots & 0\\
    \vdots & \vdots & \vdots & \ddots & \vdots \\
    \frac{h^{(n-1)}(D)}{h^{(0)}(D)} & 0 & 0 & \cdots & 1
    \end{bmatrix}.
\end{align}

To satisfy the TB condition for the feedback encoder, we use a similar procedure as that for  turbo codes \cite{Douillard2004}.
First, we use a pre-encoding operation to encode from the all-zero state and obtain a final state. Depending on that final state, an initial state is selected and the message re-encoded from this state satisfies the TB condition.  

\subsection{Constructing the Dual Trellis}
The primal trellis associated with a rate-$(n-1)/n$ ZTCC has $2^{n-1}$ outgoing branches per state. Performing SLVD over the primal trellis when $n > 2$ is highly complex. In \cite{Yang2022}, the low decoding complexity of SLVD for rate-$1/n$ convolutional codes relies on the fact that each state only has $2$ outgoing branches. In order to efficiently perform SLVD, we consider the dual trellis proposed by Yamada \emph{et al.} \cite{Yamada1983}. 

We briefly explain the dual trellis construction for parity check matrix \\ $H(D) = [h^{(n-1)}(D), h^{(n-2)}(D), \dots, h^{(0)}(D)]$. First, we define the maximum instant response order $\lambda$ as
\begin{align}
    \lambda \triangleq \max\{j\in\Brace{0,1,\dots, n-1}:  h_0^{(j)}=1\}.
\end{align}
The state of the dual trellis is represented by the partial sums of $(v+1)$ adders in the observer canonical form of $H(D)$. At time index $j$, $j=0,1,\dots, n-1$, the state is given by
\begin{align}
    \bms^{(j)} = [s_v^{(j)}, s_{v-1}^{(j)},  \dots, s_0^{(j)}].
\end{align}
Next, we show how the state $\bms^{(j)}$ evolves in terms of the output bits $\bmy_k = [y_k^{(0)}, y_k^{(1)}, \dots, y_k^{(n-1)}]$, $k=1,2,\dots, N/n$, so that a dual trellis can be established.

\begin{figure}[t]
\centering
\includegraphics[width=17.5pc]{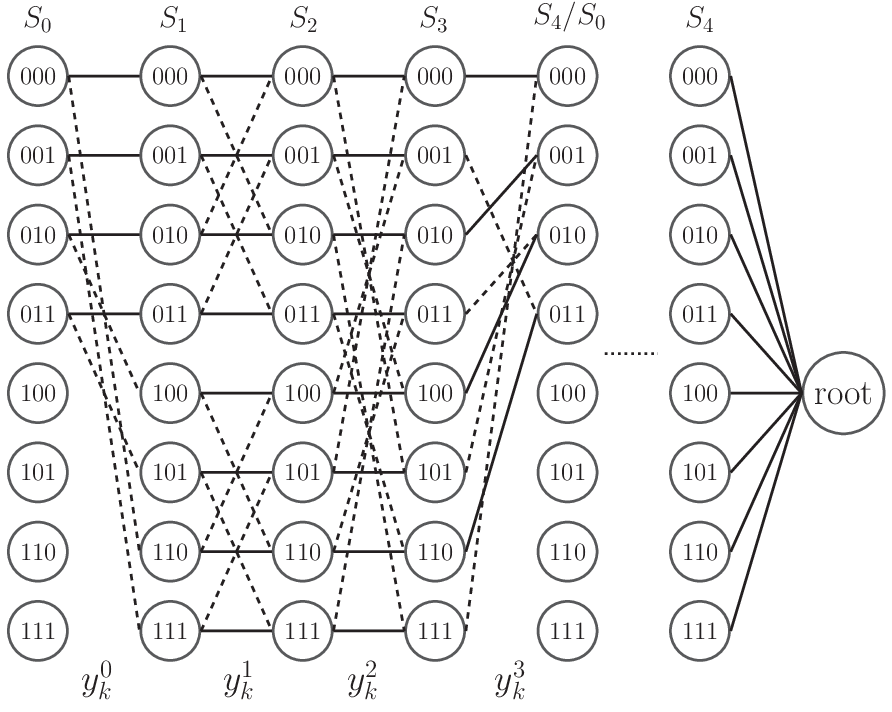}
\caption{Dual trellis diagram for rate-$3/4$ TBCC with a root node at the end for encoder $H=(7,5,2,6)$ with $v=2$. Solid lines represent $0$ paths and dashed lines represent $1$ paths.}
\label{Fig:root}
\end{figure}

\textit{Dual trellis construction for $\bmy_k = [y_k^{(0)}, y_k^{(1)}, \dots, y_k^{(n-1)}]$}:
\begin{itemize}
    \item[1)] At time $j=0$, $\bms^{(0)}=[0, s_{v-1}^{(j)}, s_{v-2}^{(j)}, \dots, s_{0}^{(j)}]$, where $s_i^{(0)}\in\{0,1\}$. Namely, only $2^v$ states exist at $j=0$.
  \item[2)] At time $j$, $j<n-1$, draw branches from each state $\bms^{(j)}$ to the states $\bms^{(j+1)}$ that satisfy
  \begin{align}
      \bms^{(j+1)} = \bms^{(j)} + y_k^{(j)}\bmh^{(j)},\quad  y_k^{(j)} \in \{0,1\} . \label{eq: transition}
  \end{align}
  \item[3)] At time $j=n-1$, draw branches from each state $\bms^{(n-1)}$ to state $\bms^{(n)}$ by
  \begin{align}
      \bms^{(n)} = \Big(\bms^{(n-1)} + y_k^{(n-1)}\bmh^{(n-1)}\Big)^r,\   y_k^{(n-1)} \in \{0,1\},
  \end{align}
  where $(a_v, a_{v-1}, \dots,a_{1}, a_0)^r = (0, a_v, a_{v-1}, \dots, a_1)$.
  \item[4)] For time $j = \lambda$, draw a branch from each state $\bms^{(\lambda)}$ according to \eqref{eq: transition} only for $y_k^{(\lambda)} = s_0^{(\lambda)}$.
\end{itemize}
After repeating the above construction for each $\bmy_k$, $k = 1, 2,\dots, N/n$, we obtain the dual trellis associated with the $(n, n-1, v)$ convolutional code. Since the primal trellis is of length $N/n$, whereas the dual trellis is of length $N$, the dual trellis can be thought of as expanding the primal trellis length by a factor of $n$, while reducing the number of outgoing branches per state from $2^{n-1}$ to less than or equal to $2$. Fig. \ref{Fig:root} illustrates the dual trellis structure for a rate-$3/4$ code with $2$ memory elements.

For an $(n, n-1, v)$ CC, zero termination over the dual trellis requires at most $n\lceil v/(n-1) \rceil$ steps. In our implementation, a breadth-first search identifies the zero-termination input and output bit patterns that provide a trajectory from each possible state $\bms$ to the zero state. The input and output bit patterns have lengths $(n-1)\lceil v/(n-1) \rceil$ and $n\lceil v/(n-1) \rceil$ respectively.

\section {Serial List Viterbi Decoding for TBCCs}
\label{sec: SLVD}
This section considers three SLVD methods that apply to both the standard and punctured high-rate TBCCs. SLVD enumerates possible paths through the trellis, starting from the lowest weight path, stopping once the first path that satisfies both the CRC and the TB condition is reached. Information about the previously investigated paths and path metrics is required to find the next optimal path. 

To efficiently implement SLVD, we use the tree-trellis algorithm (TTA) \cite{TTA1991}, which maintains a sorted list of nodes indexed by path metric. These nodes either correspond to a previously unexplored ending state in the trellis or to a previously explored path and a detour. This approach allows the efficient determination of the next path to be explored if the current one does not satisfy both the CRC and TB conditions. In order to efficiently maintain the sorted list of nodes, this paper uses a Min Heap \cite{minheap_ref}, which is easier to implement than the Red-Black tree \cite{Roder2006}\cite{Hinze99}  and has the same $\mathcal{O}(\log{} \ell)$ time complexity, where $\ell$ is the number of elements in the heap. This simpler Min Heap implementation comes at the cost of an increased memory requirement compared to the Red-Black Tree since the heap has to maintain every element, as opposed to the best $L$ elements for the tree.

\subsection {Single Trellis Decoder}
\label{subsec: single trellis}

\begin{figure}[t]
\centering
\includegraphics[width=20pc]{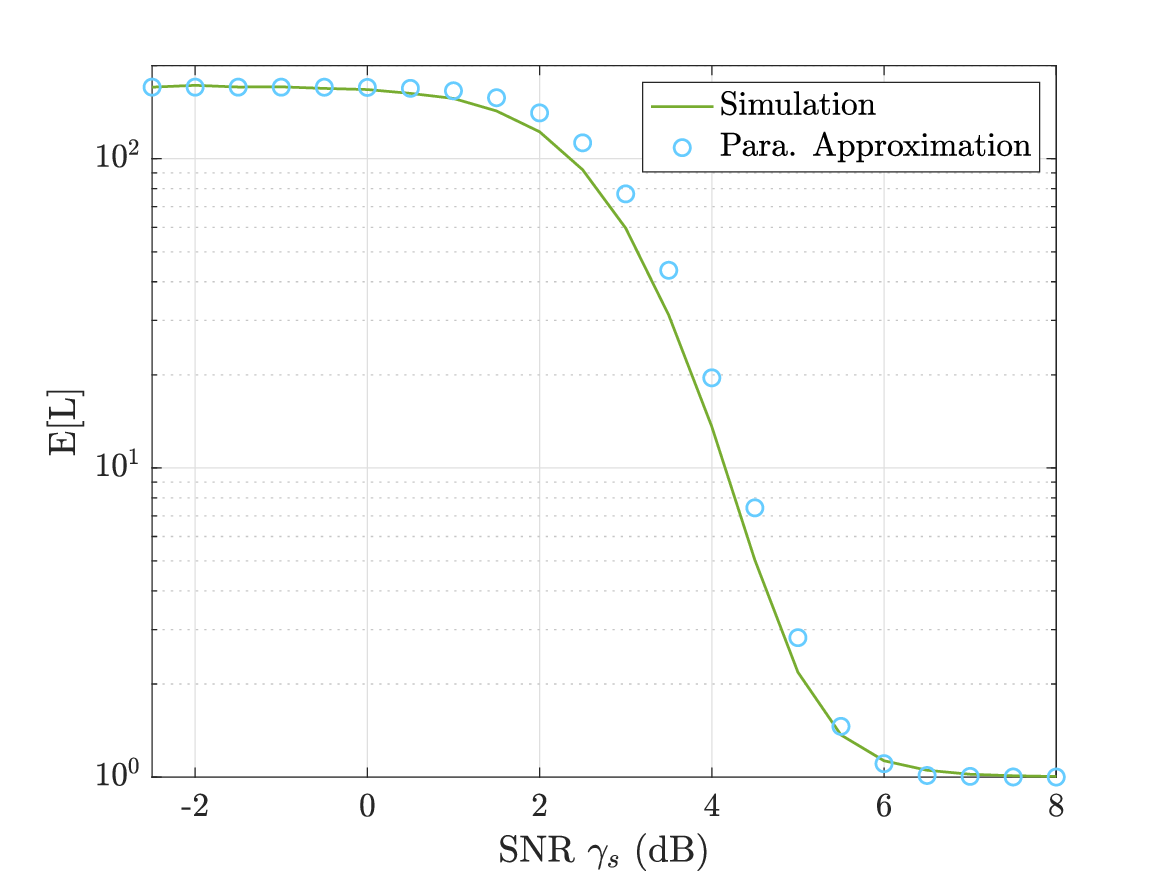}
\caption{Comparison of the parametric approximation \cite{Yang2022} of the expected list rank $\E[L]$ with the simulated results for the $v=4,m=3$ CRC-TBCC generated by $H=(33, 25, 37, 31)$ with blocklength of $128$. The optimal CRC used is 0x9. }
\label{Fig:parametric approx}
\end{figure}



To adapt SLVD to handle the multiple terminating states that are possible with a TBCC, a root node is added as shown in Fig. \ref{Fig:root}. The root node connects to all terminating states of the trellis. The Hamming distance of the branch metric for the branch connecting any state to this root node is zero. This additional root node allows the trellis to end in a single state. 

\subsection {Multi-Trellis Decoder}
\label{subsec: multi trellis}

Decoding on a single dual trellis (single-trellis approach) leads to complexity issues, 
because the decoder goes through a number of paths that pass neither the TB check nor CRC, resulting in high expected list ranks at low SNRs. To decode more efficiently, we propose a multi-trellis approach 
that includes only TB paths in the list.

The multi-trellis approach constructs $2^v$ 
trellises. 
Each multi-trellis follows the same structure as the original punctured or dual trellis, but with only one starting and ending state to enforce the TB condition. In a conceptually similar manner to how a root node was added to the dual trellis in Fig. \ref{Fig:root}, a root node is also added to the multi-trellis approach, but paths to the root node only come from the single ending state in each trellis that guarantees the TB condition.

Since all paths found using this approach will be TB, this significantly reduces the expected list size at low SNRs. However, at high SNRs, the multi-trellis approach has a substantially higher decoding complexity due to the additional upfront cost of constructing the dual trellises. In this case, the extra resources taken to initialize the multi-trellis approach bring down the overall decoder efficiency.

\subsection {Wrap-Around Viterbi Algorithm Decoder}
\label{subsec: WAVA}

As the constraint length of a TBCC increases, the number of states grows exponentially. The multi-trellis approach becomes impractical due to both time and memory for constructing the trellises. Thus, we consider a non-ML single-trellis decoder that uses WAVA \cite{Shao2003} to reduce the average list rank.
In \cite{King2023}, the authors show that a WAVA-inspired parallel list Viterbi decoder achieves good performance with low complexity.

\begin{figure}[t]
\centering
\includegraphics[width=20pc]{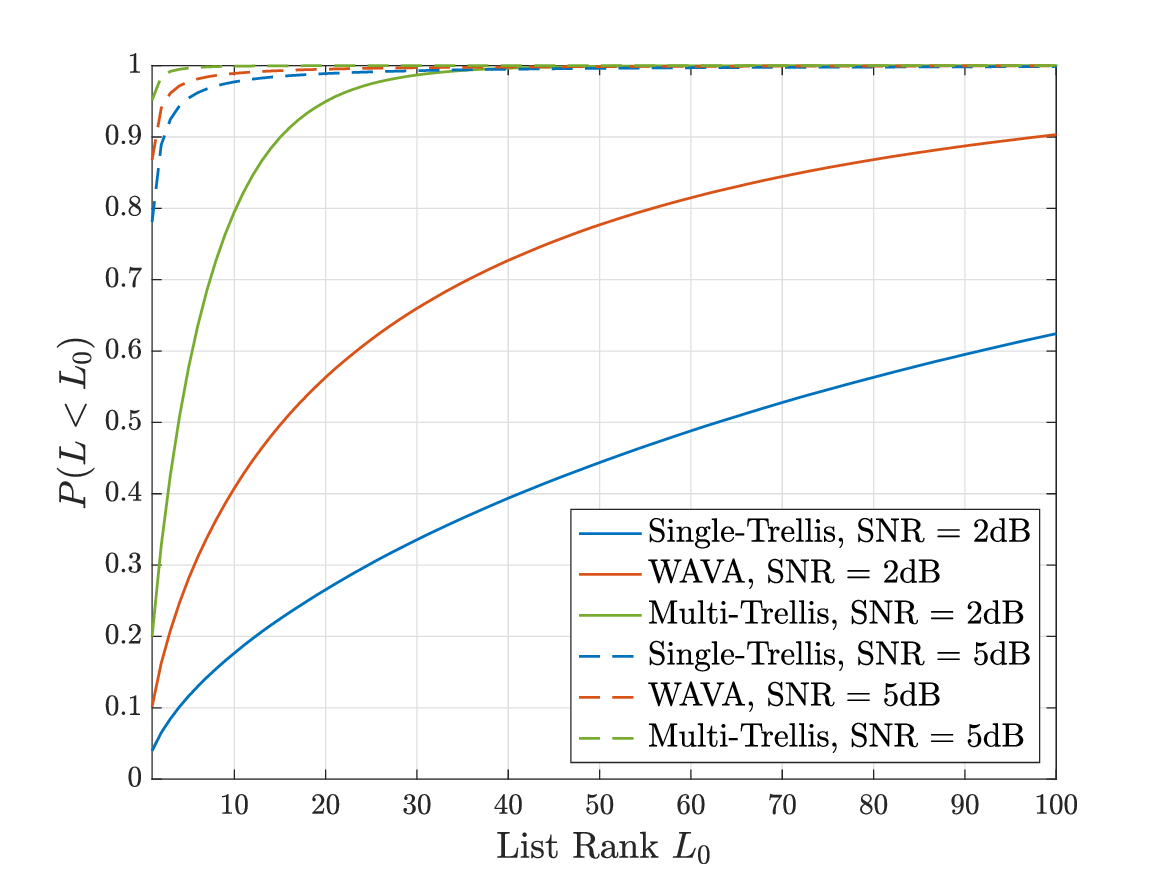}
\caption{Cumulative distribution function (CDF) of list ranks for the single-trellis, multi-trellis, and WAVA decoding approaches at SNR $\gamma_s = 2$ and $5$ dB for the $(33, 25, 37, 31)$  TBCC with blocklength of $128$.}
\label{Fig:distributions}
\end{figure}

Fig. \ref{Fig:parametric approx} extends the parametric approximation of the expected list rank shown in \cite{Yang2022} to the single-trellis decoding of high-rate CRC-TBCCs. The approximation lines up well with the simulation results for a $v=4$ TBCC with a degree-$3$ CRC. In addition, an initial examination of the list rank distributions for the three decoding schemes at SNR points of $2$ and $5$ dB is presented in Fig.  \ref{Fig:distributions}. At a low SNR, the multi-trellis approach maintains an extremely small list rank compared to the other approaches. The WAVA approach also has a substantially larger probability of a small list rank compared to the single-trellis, but not as small as the multi-trellis. Although the ordering is preserved as SNR increases, additional pre-processing to reduce list rank is inconsequential since the list ranks of all three approaches are low. Considering the extra complexity of constructing the multi-trellis approach and the non-ML nature of the WAVA approach, we use the single-trellis SLVD for simulations.
Different distributions lead to different $\E[I]$ and $\E[L]$ values when evaluating the decoding complexity, where $\E[I]$ is the expected number of insertions to maintain the sorted list of path metric differences.

The non-ML decoder with WAVA proceeds in two steps.
In the first step, the algorithm initializes each state of a single dual or punctured trellis with all zero metrics.  It then performs two iterations of add-compare-select (ACS) along the trellis. Each time the end of the trellis is encountered, the initial states of the trellis are initialized to the cumulative metrics in the final states. At the end of the first iteration, if the optimal path satisfies TB and CRC conditions, the algorithm outputs this path and stops decoding. In the second step, SLVD runs on the ending metrics of the second trellis iteration. This decoding algorithm improves the reliability of the final decision for the optimal traceback path and decreases the expected list rank while keeping the complexity low. The WAVA metrics are not ML and this algorithm 
has slightly worse decoding performance than the other two ML approaches.

\section{Optimal CRC Polynomial Design}
\label{sec: CRC design}

This section presents two approaches for designing CRC polynomials that maximize the minimum distance and minimize the number of nearest neighbors: a trellis enumeration method extended from \cite{Yang2022} and a list decoding sieve method proposed in \cite{WeselISTC2023}. As a case study, this paper mainly focuses on the rate-$3/4$ systematic feedback convolutional codes in \cite[Table 12.1(e)]{LinCostello2004} and the punctured rate-$3/4$ convolutional codes in \cite{HACCOUN1989}. 


\subsection{CRC Polynomials for Standard Zero-Terminated Codes}\label{subsec: DSO CRC ZTCC feedback}

In this paper, we focus on the low FER regime. Thus, the 
CRC polynomials identified in this paper simply maximize the minimum distance $d_{\min}$ of the concatenated code and minimizes the number of nearest neighbors. Examples in \cite{Yang2022} indicate that 
CRC polynomials designed in this way can provide optimal or near-optimal performance for a wide range of SNRs.

We apply the CRC polynomial design algorithm in \cite{Yang2022} to identify CRC polynomials for high-rate ZTCCs. The first step is to collect the irreducible error events (IEEs), which are ZT paths on the trellis that deviate from the zero state once and rejoin it once. IEEs with a very large output Hamming weight do not affect the choice of optimal CRC polynomials.  In order to reduce the runtime of the CRC optimization algorithm, IEEs with output Hamming weight greater than or equal to a threshold $\tilde{d}$ are not considered. Dynamic programming constructs all ZT paths of length equal to $N/n$ and output weight less than $\tilde{d}$. Finally, we use the resulting set of ZT paths to identify the degree-$m$ 
optimal CRC polynomial for the rate-$(n-1)/n$ CC.

In \cite{Yang2022}, Yang \emph{et al.} provided a useful result for selecting threshold $\tilde{d}$.
\begin{theorem}[Th. 2, \cite{Yang2022}]\label{theorem: upper bound}
    Define the higher-rate code $\C_h$ by
    \begin{align}
        \C_{h} \triangleq \big\{\bm{c}\in\{0,1\}^n: \bm{c} = \bm{v}\bm{G}, \forall \bm{v}\in\{0,1\}^{k+m} \big\},
    \end{align}
where $\bm{G}\in\{0,1\}^{(k+m)\times n}$ is the matrix representation of the convolutional encoder. Given a specified CRC degree $m$ and a higher-rate code $\C_h$ with distance spectrum $B_{d_{\min}^h}, \dots, B_n$, define $w^*$ as the minimum $w$ for which $\sum_{d=d_{\min}^h}^w B_d\ge 2^m$. For any degree-$m$ CRC polynomial, we have $d_{\min}^l\le 2w^*$.
\end{theorem}

Theorem \ref{theorem: upper bound} shows that it suffices to choose $\tilde{d} = 2w^* + 1$ to identify the degree-$m$ 
CRC polynomial that maximizes the minimal distance. In practice, the weight $w^*$ can be efficiently determined from the weight enumerating function of a convolutional code \cite[p. 488]{LinCostello2004}.

 \begin{table}[t] 
\centering
\caption{
Optimal CRC Polynomials for Standard Rate-$3/4$ ZTCC at Blocklength $N=128$ Generated by $H=(33, 25, 37, 31)$ With $v=4$, by $H=(47, 73, 57, 75)$ With $v=5$, \\
and by $H=(107, 135, 133, 141)$ With $v=6$}
\label{tab:ZT_codes_with_CRC_feedback}
\begin{tabular}{P{0.1cm} P{0.1cm} P{0.5cm} P{0.3cm} P{0.4cm} P{0.5cm} P{0.3cm} P{0.4cm} P{0.5cm} P{0.3cm} P{0.4cm}}
\hline
\clineB{1-11}{1.2}
& \\[\dimexpr-\normalbaselineskip+2pt]
\multicolumn{1}{c}{$K$} & \multicolumn{1}{c}{$m$} & \multicolumn{3}{c}{$v=4$} & \multicolumn{3}{c}{$v=5$} & \multicolumn{3}{c}{$v=6$} \\ 
\multicolumn{1}{c}{} & \multicolumn{1}{c}{} & CRC  & $d_{\min}$ & $A_{d_{\min}}$  & CRC  & $d_{\min}$  & $A_{d_{\min}}$  &CRC  & $d_{\min}$ & $A_{d_{\min}}$ \\
 \noalign{\vskip 0.5mm}    
 \hline 
$90$ &$0$ &  0x1 & 4 & 60 &  0x1 & 5 & 200 & 0x1 & 6 & 736\\
$89$ &$1$ &  0x3 & 4 & 30 & 0x3 & 5 & 113 & 0x3 & 6 & 331\\
$88$ &$2$ & 0x7 & 5 & 85 & 0x7 & 5 & 56 & 0x7 & 6 & 106\\
$87$ &$3$ & 0x9 & 5 & 1 & 0x9 & 5 & 1 & 0xB & 6 & 34\\
$86$ &$4$  & 0x1B & 6 & 251 & 0x15 & 6 & 54 & 0x1D & 6 & 3\\
$85$ &$5$ &  0x25 & 6 & 32 & 0x25 & 7 & 156 & 0x25 & 7 & 27\\
$84$ &$6$ & 0x4D & 7 & 155 & 0x7B & 7 & 76 & 0x6F & 7 & 1\\
$83$ &$7$ &  0xF3 & 7 & 45 & 0xED & 8 & 194 & 0x97 & 8 & 12\\
$82$ &$8$ &  0x1E9 & 8 & 145 & 0x1B7 & 8 & 25 & 0x1B5 & 9 & 375\\ 
$81$ &$9$ &  0x31B & 8 & 27 & 0x3F1 & 8 & 1 & 0x2F1 & 9 & 65\\
$80$ &$10$ &  0x5C9 & 9 & 168 & 0x66F & 9 & 2 & 0x59F & 10 & 490\\
$79$ &$11$ &  0xC2B & 10 & 1015 & 0xE8D & 10 & 293 & 0xD2D & 10 & 42 \\
 \hline
\clineB{1-11}{1.2}
\end{tabular}
\end{table}

Table \ref{tab:ZT_codes_with_CRC_feedback} presents the
optimal CRC polynomials for ZTCCs generated with $H=(33, 25, 37, 31)$, $H=(47, 73, 57, 75)$, and $H=(107, 135, 133, 141)$. Table \ref{tab:ZT_codes_with_CRC_feedback} also shows the minimum distance $d_{\min}$ and number of nearest neighbors $A_{d_{\min}}$ of the CRC-ZTCCs.
The design assumes a fixed blocklength $N = 128$ bits. Due to the overhead caused by the CRC bits and by zero termination, the rates of CRC-ZTCCs are less than $3/4$. Specifically, for a given information length $K$, CRC degree $m$ and an $(n, n-1, v)$ encoder, the blocklength $N$ for a CRC-ZTCC is given by
\begin{align}
    N = \left (K+m + (n-1) \left \lceil \frac{v}{n-1} \right \rceil \right ) \frac{n}{n-1}, \label{eq: blocklength ZTCC}
\end{align}
giving
\begin{align}
    R= \frac{K}{N} = \frac{n-1}{n}\frac{K}{K+m+(n-1)\lceil \frac{v}{n-1} \rceil }.
\end{align}
 We see from \eqref{eq: blocklength ZTCC} that the $(n, n-1, v)$ convolutional encoder can accept any CRC  degree $m$ as long as $K+m$ is divisible by $(n-1)$.

\subsection{CRC Polynomials for Punctured Zero-Terminated Codes}\label{subsec: DSO CRC ZTCC punctured}

In \cite{WeselISTC2023}, the authors proposed an efficient list decoding sieve method to identify the distance-optimal CRC polynomial of a given degree. This approach takes a noiseless all-zeros codeword as the received signal and performs serial list Viterbi decoding to
explore codewords in order of increasing Hamming weight.
For each new codeword added to the list, we check if it passes any of the degree-$m$ CRC polynomials. If a CRC polynomial can eliminate all codewords of a certain Hamming weight, the sieve approach keeps this CRC polynomial and codewords of the next greater weight are explored. The list decoding sieve continues until it reaches a codeword weight where all CRC polynomials check at least one codeword. This weight is the largest $d_{\min}$ that a degree-$m$ CRC polynomial can achieve. The CRC polynomial that checks the least number of codewords at $d_{\min}$ is selected as the optimal CRC polynomial. This approach is computationally more efficient than the error event construction method of Yang \cite{Yang2022} while producing the same results. 

We follow the puncturing patterns provided \cite{HACCOUN1989} for rate-$3/4$ CCs with $v=4,5,6$. These punctured CCs are obtained from puncturing $4$ out of every $6$ bits ($3$ symbols) for rate-$1/2$ convolutional codes. The blocklength of all CRC-CCs is $N=128$ bits.

\begin{table}[t]
\centering
\caption{
Optimal CRC Polynomials for Punctured rate-$3/4$ ZTCC at Blocklength $N=128$ Generated by $G=(23,25)$ With $v=4$, 
by $G=(53,75)$ With $v=5$, 
and by $G=(133,171)$ With $v=6$}
\label{tab:ZT_codes_with_CRC_punc}
\begin{tabular}{P{1cm} P{0.1cm}  P{0.4cm} P{0.3cm} P{0.5cm} P{0.4cm} P{0.3cm} P{0.5cm}  P{0.4cm} P{0.3cm} P{0.5cm}}
\hline
\clineB{1-11}{1.2}
& \\[\dimexpr-\normalbaselineskip+2pt]
$K+v$ & $m$ & \multicolumn{3}{c}{$v=4$} & \multicolumn{3}{c}{$v=5$} & \multicolumn{3}{c}{$v=6$} \\ 
 & & CRC  & $d_{\min}$ & $A_{d_{\min}}$ & CRC  & $d_{\min}$  & $A_{d_{\min}}$ &CRC  & $d_{\min}$ & $A_{d_{\min}}$ \\ \noalign{\vskip 0.5mm}  
 \hline 
$96$ &$0$  & 0x1 & 3 & 31 & 0x1 & 4 & 29 & 0x1 & 5 & 223 \\
$95$ &$1$  & 0x3 & 4 & 29 & 0x3 & 5 & 224 & 0x3 & 5 & 112 \\
$94$ &$2$  & 0x7 & 6 & 2173 & 0x7 & 5 & 83 & 0x7 & 6 & 427 \\
$93$ &$3$  & 0xF & 6 & 597 & 0x9 & 6 & 379 & 0x9 & 6 & 135 \\
$92$ &$4$  & 0x11 & 6 & 323 & 0x1F & 6 & 53 & 0x13 & 7 & 245 \\
$91$ &$5$  & 0x27 & 6 & 101 & 0x39 & 7 & 213 & 0x23 & 8 & 1206 \\
$90$ &$6$  & 0x71 & 7 & 286 & 0x79 & 7 & 46 & 0x65 & 8 & 590 \\
$89$ &$7$  & 0xC7 & 7 & 54 & 0x85 & 8 & 216 & 0xFD & 8 & 122 \\
$88$ &$8$  & 0x199 & 8 & 407 & 0x153 & 8 & 22 & 0x163 & 8 & 17 \\
$87$ &$9$  & 0x20B & 8 & 68 & 0x353 & 9 & 247 & 0x247 & 10 & 2158 \\
$86$ &$10$  & 0x439 & 9 & 400 & 0x7CD & 10 & 1631 & 0x4E7 & 10 & 342 \\
 \hline
\clineB{1-11}{1.2}
\end{tabular}
\end{table}

Table \ref{tab:ZT_codes_with_CRC_punc} shows the 
optimal CRC polynomials obtained by the list decoding sieve approach for punctured ZTCCs generated with $G=(23,25)$, $G=(53,75)$, $G=(133,171)$, as well as the corresponding $d_{\min}$ and $A_{d_{\min}}$ of the CRC-ZTCCs. For a feedforward code, the ZT condition is satisfied by inputting $v$ zero bits at the end of the information sequence. Thus, for a given information length $K$, the  blocklength $N$ of the punctured code is given by 
\begin{align}
    N = \left (K+m + v \right ) \frac{n}{n-1}. \label{eq: blocklength ZTCC punctured}
\end{align}

The real rate of this code is given by
\begin{align}
    R= \frac{K}{N} = \frac{n-1}{n}\frac{K}{K+m+v }.
\end{align}
In general, the punctured CRC-ZTCCs have smaller $d_{\min}$ values and more nearest neighbors than the standard CRC-ZTCCs.

\subsection{CRC Polynomials for Standard Tail-Biting Codes}
\label{subsec: DSO CRC TBCC feedback}

The design of 
CRC polynomials for standard high-rate TBCCs follows the two-phase design algorithm shown in \cite{Yang2022}. This algorithm is briefly explained below.

Consider a TB trellis $T = (V, E, \A)$ of length $N$, where $\A$ denotes the set of output alphabet, $V$ denotes the set of states, and $E$ denotes the set of edges described in an ordered triple $(s, a, s')$ with $s, s'\in V$ and $a\in\A$ \cite{Koetter2003}. Assume $|V| = 2^v$ and let $V_0 = \{0, 1, \dots, 2^v-1\}$. Define the set of IEEs at state $\sigma\in V$ as
\begin{align}
    \IEE(\sigma)\triangleq\bigcup_{l=1,2,\dots,N}\overline{\IEE}(\sigma,l),
\end{align}
where
\begin{equation}
    \begin{aligned}
         \overline\IEE(\sigma,l) \triangleq&\{(\bm{s},\bm{a})\in V_0^{l+1}\times \A^{l}: s_0=s_l=\sigma; \ \\
         &\forall j, 0<j<l,\ s_{j}\notin\{0, 1,\dots,\sigma\}\}.
    \end{aligned}
\end{equation}

By concatenating elements in $\IEE(\sigma)$, one can build an arbitrarily long TB path that starts and ends at state $\sigma$.
The first phase is called the collection phase, during which the algorithm collects $\IEE(\sigma)$ with output Hamming weight less than the threshold $\tilde{d}$ over a sufficiently long TB trellis. The second phase is called the search phase, during which the algorithm first reconstructs all TB paths of length $N/n$ and output weight less than $\tilde{d}$ via concatenation of the IEEs and circular shifting of the resulting path. Then, using these TB paths, the algorithm searches for the degree-$m$ optimal CRC polynomial by maximizing the minimum distance of the undetected TB path.

\begin{table}[t]
\centering
\caption{
Optimal CRC Polynomials for Standard rate-$3/4$ TBCC at Blocklength $N=128$ Generated by $H=(33, 25, 37, 31)$ With $v=4$, by $H=(47, 73, 57, 75)$ With $v=5$, \\
and by $H=(107, 135, 133, 141)$ With $v=6$}
\label{tab:TB_codes_with_CRC_feedback}
\begin{tabular}{P{0.1cm} P{0.1cm}  P{0.5cm} P{0.5cm} P{0.5cm} P{0.4cm} P{0.5cm} P{0.5cm}  P{0.5cm} P{0.4cm} P{0.5cm}}
\hline
\clineB{1-11}{1.2}
& \\[\dimexpr-\normalbaselineskip+2pt]
$K$ & $m$ & \multicolumn{3}{c}{$v=4$} & \multicolumn{3}{c}{$v=5$} & \multicolumn{3}{c}{$v=6$} \\ 
 & & CRC  & $d_{\min}$ & $A_{d_{\min}}$ & CRC  & $d_{\min}$  & $A_{d_{\min}}$ &CRC  & $d_{\min}$ & $A_{d_{\min}}$ \\ \noalign{\vskip 0.5mm}  
 \hline 
$96$ &$0$  & 0x1 & 4 & 64 & 0x1 & 5 & 224 & 0x1 & 6 & 864 \\
$95$ &$1$  & 0x3 & 4 & 32 & 0x3 & 5 & 128 & 0x3 & 6 & 384 \\
$94$ &$2$  & 0x7 & 5 & 96 & 0x7 & 5 & 64 & 0x7 & 6 & 128\\
$93$ &$3$  & 0x9 & 6 & 736 & 0x9 & 6 & 192 & 0xB & 6 & 36\\
$92$ &$4$  & 0x1B & 6 & 320 & 0x15 & 6 & 64 & 0x1D & 6 & 6 \\
$91$ &$5$  & 0x25 & 6 & 31 & 0x37 & 6 & 2 & 0x23 & 7 & 49 \\
$90$ &$6$  & 0x4D & 6 & 1 & 0x4F & 7 & 98 & 0x53 & 8 & 326 \\
$89$ &$7$  & 0xA3 & 7 & 70 & 0xD1 & 8 & 446 & 0xB1 & 8 & 76\\
$88$ &$8$  & 0x10D & 8 & 411 & 0x149 & 8 & 73 & 0x1D3 & 8 & 8 \\ 
$87$ &$9$  & 0x2ED & 8 & 138 & 0x255 & 8 & 14 & 0x3F7 & 9 & 208 \\
$86$ &$10$  & 0x63B & 8 & 23 & 0x70F & 8 & 1 &  0x529 & 9 & 90 \\
$85$ &$11$  & 0xCA5 & 9 & 125 & 0xD57 & 9 & 17 &  0x9BD & 10 & 387 \\
$84$ &$12$  & 0x1ED7 & 10 & 904 & 0x1B41 & 10 & 339 &  0x10AF & 10  & 53 \\
 \hline
\clineB{1-11}{1.2}
\end{tabular}
\end{table}

Table \ref{tab:TB_codes_with_CRC_feedback} presents the optimal CRC polynomials for TBCCs generated with $H=(33, 25, 37, 31)$, $H=(47, 73, 57, 75)$, and $H=(107, 135, 133, 141)$. \footnote{This table is updated from \cite{Sui2022}, as an error was discovered in the previous CRC polynomial design procedure.} The design assumes a fixed blocklength $N = 128$.  TB encoding avoids the rate loss caused by the overhead of zero termination. Specifically, for a given information length $K$, CRC degree $m$ and an $(n, n-1, v)$ encoder, the blocklength $N$ for a CRC-TBCC is given by
\begin{align}
     N = \left( {K+m} \right ) \frac{n}{n-1} \label{eq: blocklength},
\end{align}
giving
\begin{align}
    R = \frac{K}{N} = \frac{n-1}{n}\frac{K}{K+m}.
\end{align}

\begin{figure}[t]
\centering
\includegraphics[width=20pc]{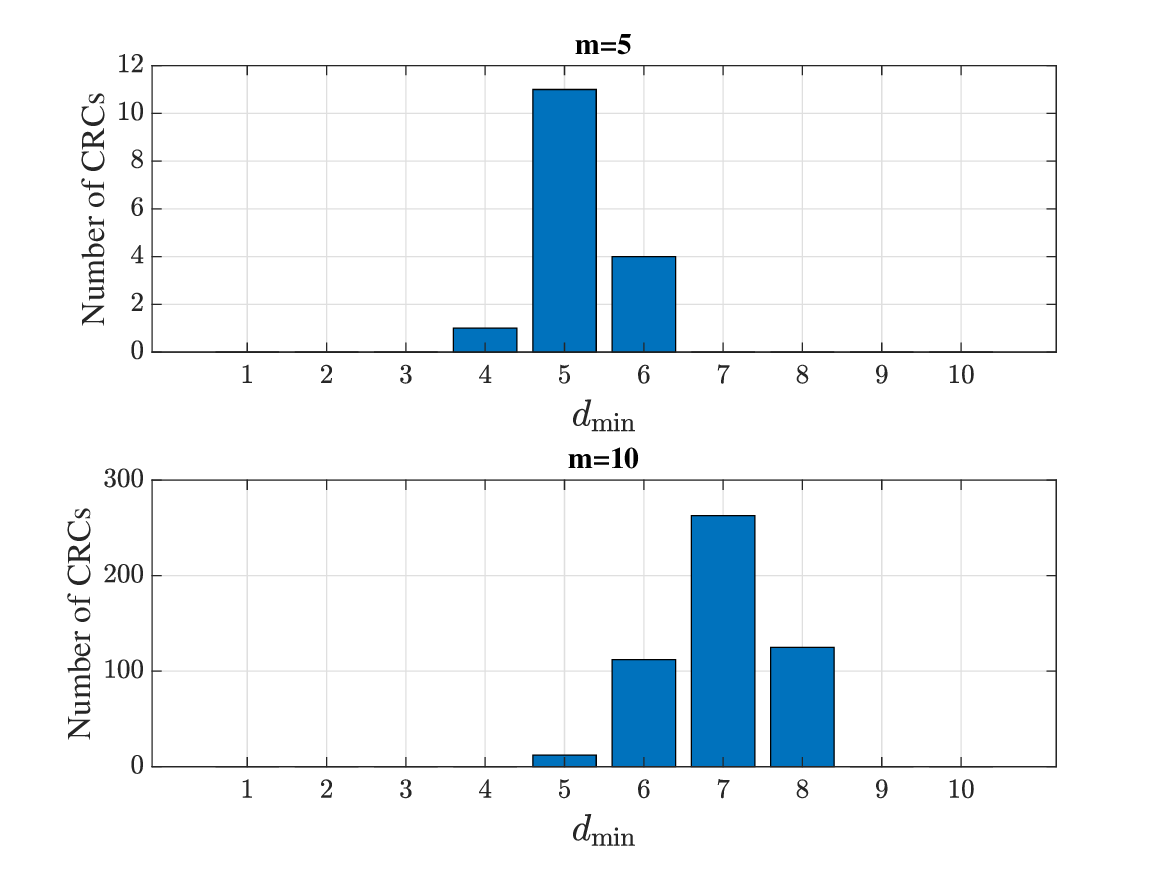}
\caption{Distribution of $d_{\min}$ of all CRC polynomials of degree $5$ (top) and $10$ (bottom) when concatenated with the $v=4$ standard TBCC generated b $H=(33,25,37,31)$.}
\label{Fig:dmin}
\end{figure}

Fig. \ref{Fig:dmin} shows the distribution of minimum distances for all degree-$5$ and $10$  CRC polynomials for $v=4$ CRC-TBCCs. While multiple CRC polynomials have the same maximized $d_{\min}$, their $A_{d_{\min}}$ differ. The optimal CRC polynomial designs have the minimal $A_{d_{\min}}$ value. This range of $d_{\min}$ values validates the effectiveness of our CRC polynomial design approach in the high-rate scenario.

 \subsection{CRC Polynomials for Punctured Tail-Biting Codes}
 \label{subsec: DSO CRC TBCC punctured}

The sieve method described in Sec. \ref{subsec: DSO CRC ZTCC punctured} is extended to TBCCs efficiently with the application of the multi-trellis SLVD, since all codewords discovered by the multi-trellis satisfy the tail-biting condition. Table \ref{tab:TB_codes_with_CRC_punc} shows the optimal CRC polynomial designs for $v=4,5,6$ punctured TBCCs generated with $G=(23,25)$, $G=(53,75)$, $G=(133,171)$. Since there is no overhead of terminations, the punctured TBCCs have the same blocklength and rate at a given information length. 
Similar to the ZT case, a rate-$1/2$ code punctured to rate-$3/4$ has slightly worse $d_{\min}$ and $A_{d_{\min}}$ values compared to a standard rate-$3/4$ TBCC.  

\begin{table}[t]
\centering
\caption{
Optimal CRC Polynomials for Punctured rate-$3/4$ TBCC at Blocklength $N=128$ Generated by $G=(23,25)$ With $v=4$,
by $G=(53,75)$ With $v=5$,
and by $G=(133,171)$ With $v=6$}
\label{tab:TB_codes_with_CRC_punc}
\begin{tabular}{P{0.1cm} P{0.1cm}  P{0.4cm} P{0.3cm} P{0.5cm} P{0.4cm} P{0.3cm} P{0.5cm}  P{0.4cm} P{0.3cm} P{0.5cm}}
\hline
\clineB{1-11}{1.2}
& \\[\dimexpr-\normalbaselineskip+2pt]
$K$ & $m$ & \multicolumn{3}{c}{$v=4$} & \multicolumn{3}{c}{$v=5$} & \multicolumn{3}{c}{$v=6$} \\ 
 & & CRC  & $d_{\min}$ & $A_{d_{\min}}$ & CRC  & $d_{\min}$  & $A_{d_{\min}}$ &CRC  & $d_{\min}$ & $A_{d_{\min}}$ \\ \noalign{\vskip 0.5mm}  
 \hline 
$96$ &$0$  & 0x1 & 3 & 32 & 0x1 & 4 & 32 & 0x1 & 5 & 256 \\
$95$ &$1$  & 0x3 & 4 & 32 & 0x3 & 5 & 256 & 0x3 & 5 & 128 \\
$94$ &$2$  & 0x7 & 6 & 2512 & 0x7 & 5 & 96 & 0x7 & 6 & 512 \\
$93$ &$3$  & 0xF & 6 & 688 & 0x9 & 6 & 448 & 0x9 & 6 & 160 \\
$92$ &$4$  & 0x11 & 6 & 368 & 0x15 & 6 & 96 & 0x1B & 6 & 64 \\
$91$ &$5$  & 0x33 & 6 & 176 & 0x25 & 6 & 7 & 0x3F & 8 & 1637 \\
$90$ &$6$  & 0x71 & 6 & 7 & 0x55 & 7 & 224 & 0x77 & 8 & 767 \\
$89$ &$7$  & 0xD5 & 6 & 2 & 0xC3 & 8 & 1166 & 0xBD & 8 & 365 \\
$88$ &$8$  & 0x1EB & 7 & 20 & 0x129 & 8 & 281 & 0x101 & 8 & 27 \\
$87$ &$9$  & 0x343 & 8 & 211 & 0x367 & 8 & 79 & 0x2B7 & 8 & 4 \\
$86$ &$10$  & 0x677 & 8 & 69 & 0x41D & 8 & 4 & 0x40D & 8 & 1 \\
 \hline
\clineB{1-11}{1.2}
\end{tabular}
\end{table}

\section{Complexity Analysis}
\label{sec: complexity analysis}

In this section, we will discuss the decoding complexity for all SLVD methods presented in Sec. \ref{sec: SLVD}. Section \ref{subsec: feedback complexity} covers the complexity analysis of rate-$(n-1)/n$ standard ZTCCs and TBCCs on a dual trellis. Section \ref{subsec: punctured complexity} provides the decoding complexity equations for punctured CCs. Finally, section \ref{subsec: complexity comparison} visualizes the performance-complexity trade-offs between the standard and punctured codes. The WAVA decoder is a low-complexity alternative for TBCCs, and we explore its complexity and performance for standard TBCCs.

\subsection{Dual Trellis SLVD for Standard ZTCC and TBCC}
\label{subsec: feedback complexity}

In \cite{Yang2022}, the authors provided the complexity expression for SLVD of CRC-ZTCCs and CRC-TBCCs, where the convolutional encoder is of rate $1/n$. Observe that the dual trellis has no more than $2$ outgoing branches per state, similar to the trellis of a rate-$1/n$ CC. Thus, we directly apply their complexity expression to SLVD over the dual trellis.

As noted in \cite{Yang2022}, the overall average complexity of SLVD can be decomposed into three components:
\begin{align}
    C_{\SLVD} = C_{\SSV} + C_{\trace} + C_{\List},
    \label{eq: c_slvd}
\end{align}
where $C_{\SSV}$ denotes the complexity of a standard soft Viterbi (SSV), $C_{\trace}$ denotes the complexity of the \emph{additional} traceback operations required by SLVD, and $C_{\List}$ denotes the average complexity of inserting new elements to maintain an ordered list of path metric differences.

$C_{\SSV}$ is the complexity of ACS operations and the initial traceback operation. For CRC-ZTCCs,
\begin{align}
    C_{\SSV}&=(2^{v+1}-2)+1.5(2^{v+1}-2)+1.5(K+m-v)2^{v+1}\notag\\
  &\phantom{=}+c_1[2(K+m+v)+1.5(K+m)]. \label{eq: C_SSV}
\end{align}
For CRC-TBCCs decoded using the single-trellis, this quantity is given by
\begin{align}
    C_{\SSV}&=1.5(K+m)2^{v+1}+2^{v}+3.5c_1(K+m). \label{eq: C_SSV for CRC-TBCC 1 trellis}
\end{align}
For CRC-TBCCs with the multi-trellis approach,
\begin{align}
    C_{\SSV}&=2^v[1.5(K+m)2^{v+1}]+3.5c_1(K+m). \label{eq: C_SSV for CRC-TBCC multi-trellis}
\end{align}
The second component $C_{\trace}$ for CRC-ZTCC is given by
\begin{align}
    C_{\trace} = c_1(\E[L]-1)[2(K+m+v)+1.5(K+m)]. \label{eq: C_trace}
\end{align}
For CRC-TBCCs, $C_{\trace}$ is given by
\begin{align}
    C_{\trace} = 3.5c_1(\E[L]-1)(K+m), \label{eq: C_trace for CRC-TBCC}
\end{align}
for both single-trellis and multi-trellis approaches.

The third component, which is identical for ZT and TB, is
\begin{align}
    C_{\List} = c_2\E[I]\log(\E[I]). \label{eq: C_list}
\end{align}
For CRC-ZTCCs,
\begin{align}
  \E[I]&\le (K+m)\E[L], \label{eq: EL ZTCC}
\end{align}
and for CRC-TBCCs with either single-trellis or multi-trellis approach,
\begin{align}
  \E[I]&\le (K+m)\E[L]+2^{v}-1. \label{eq: EL TBCC}
\end{align}

In the above expressions, $c_1$ and $c_2$ are two computer-specific constants that characterize implementation-specific differences in the implemented complexity of traceback and list insertion (respectively) as compared to the ACS operations of Viterbi decoding. In this paper, we assume that $c_1 = c_2 = 1$ and use \eqref{eq: EL ZTCC} and \eqref{eq: EL TBCC} to estimate $\E[I]$ for CRC-ZTCCs and CRC-TBCCs.

\begin{figure}[t]
\centering
\includegraphics[width=20pc]{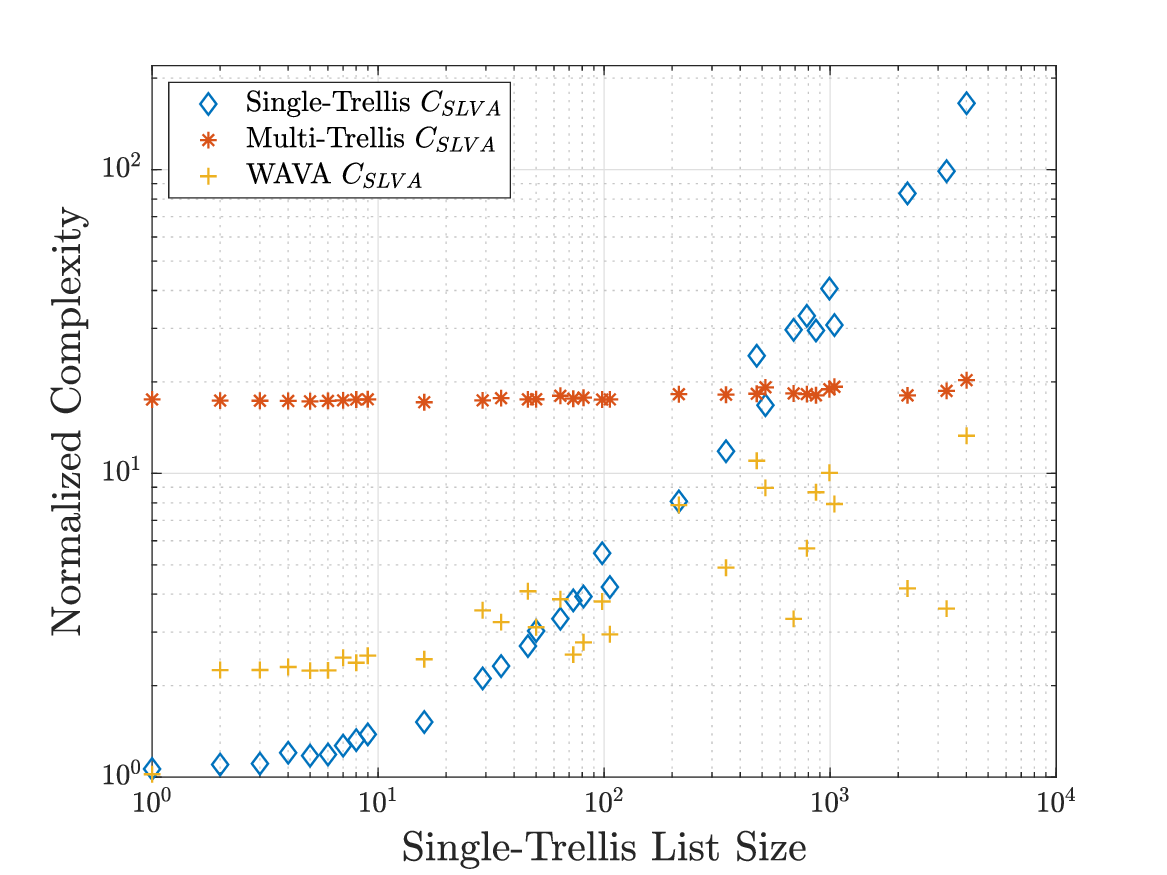}
\caption{The overall complexity comparison of the single-trellis, multi-trellis, and WAVA decoders for the TBCC generated with the $(4,3,4)$ encoder $H = (33,25,37)$, with blocklength of $128$. The CRC polynomial of degree $3$ is 0x9. All complexity values are normalized with respect to the single-trellis $C_{\SSV}$ at different list sizes.
}
\label{Fig:1trellis_ntrellis_compare}
\end{figure}

Note that $\E[I]$ and $\E[L]$ values vary depending on whether the single-trellis or multi-trellis approach is used. Using the multi-trellis approach significantly  reduces $C_{\trace}$ and $C_{\List}$ because only TB paths are included. On the other hand, as seen from \eqref{eq: C_SSV for CRC-TBCC 1 trellis} and \eqref{eq: C_SSV for CRC-TBCC multi-trellis}, the multi-trellis approach amplifies the first component $C_{\SSV}$ by nearly $2^v$. The overall trade-off is depicted in Fig. \ref{Fig:1trellis_ntrellis_compare}, which shows the complexity comparison of the three proposed SLVD methods for a $v=4, m=3$ standard CRC-TBCC decoded using the dual trellis. Random codewords with blocklength $N=128$ are generated and their single-trellis list sizes are measured by passing through a single-trellis SLVD.
The runtime of each complexity component is normalized with respect to the value of single-trellis $C_{\SSV}$. When the single-trellis list size is $1$, the multi-trellis SLVD has an overall runtime that is over $10$ times greater than that of the single-trellis SLVD.
At low noise levels, the list size of a single trellis is almost always $1$, resulting in a substantially lower runtime compared to that of a multi-trellis. As SNR decreases, there is
an exponential growth in the complexity terms $C_{\trace}$ and $C_{\List}$ for the single-trellis decoder.  The list size grows much more slowly for the mutli-trellis decoder because it does not include non-TB codewords in the list.
As a result, trellis construction is the main contributor to the complexity of multi-trellis. Thus the multi-trellis decoder has similar complexity
across all SNR levels. At a single-trellis list size of around $5\times 10^2$, the overall runtime $C_{\SLVD}$ of both approaches becomes the same. This indicates that at high SNRs, single-trellis is the optimal approach.
But when the noise level is high, the multi-trellis approach has a more favorable runtime since it guarantees to satisfy the TB condition.

Upon applying WAVA, the overall average complexity for CRC-TBCC is incremented by ACS operations during the additional forward pass, if needed. Let the probability that the optimal path of the initial traceback does not satisfy either TB or CRC condition be $P_{\WAVA}$. The list rank of the decoder is $1$ with a probability of $1-P_{\WAVA}$. Thus we have the updated complexity:
\begin{align}
    C_{\SLVD} = C_{\SSV} + P_{\WAVA}(C_{\WAVA} + C_{\trace} + C_{\List}),
\end{align}
where
\begin{align}
    C_{\WAVA}&=1.5(K+m)2^{v+1}+2^{v}.
\end{align}

The yellow data points in Fig.  \ref{Fig:1trellis_ntrellis_compare} represent the overall complexity of the WAVA decoder normalized with respect to the single-trellis $C_{\SSV}$. The complexity for initializing the WAVA decoder is about $2$ times of that for the single-trellis decoder, giving it a disadvantage at low SNRs. When list size is $1$,  the WAVA decoder matches the complexity of the single-trellis decoder since one iteration is sufficient.
At a list size of around $50$, the overall complexity of the WAVA decoder reaches the same level as the single-trellis decoder. The WAVA decoder always operates at a complexity lower than the multi-trellis decoder.

\subsection{Primal Trellis SLVD for Punctured ZTCC and TBCC}
\label{subsec: punctured complexity}

A punctured convolutional code of rate $(n-1)/n$ is obtained from puncturing the outputs of a rate-$1/2$ code. Therefore, the complexity of the SLVD for the punctured and original codes are the same, which is presented in \cite{Yang2022}. To keep this section self-contained, we will show the rate-$1/n$ complexity analysis here. 

The overall complexity of the punctured SLVD consists of the same three components as that of the standard SLVD in \ref{eq: c_slvd}.

For CRC-ZTCCs, 
\begin{equation}
 \begin{aligned}
    C_{\SSV}&=5(2^v - 1) + 3(K+m-v)2^v \\
    &+ c_1[2(K+m+v)+1.5(K+m)]. \label{eq: C_SSV_ZT_punc}
 \end{aligned}
\end{equation}
    
For CRC-TBCCs, 
\begin{align}
    C_{\SSV}&=(3K+3m+1) 2^v + 3.5c_1(K+m). \label{eq: C_SSV_TB_punc}
\end{align}

The other two components $C_{\trace}$ and $C_{\List}$, as well as the expected number of insertions $\E[I]$, remain the same for punctured SLVD as the dual-trellis SLVD (Eq. \ref{eq: C_trace} - \ref{eq: EL TBCC}).

\begin{figure}[t]
\centering
\includegraphics[width=20pc]{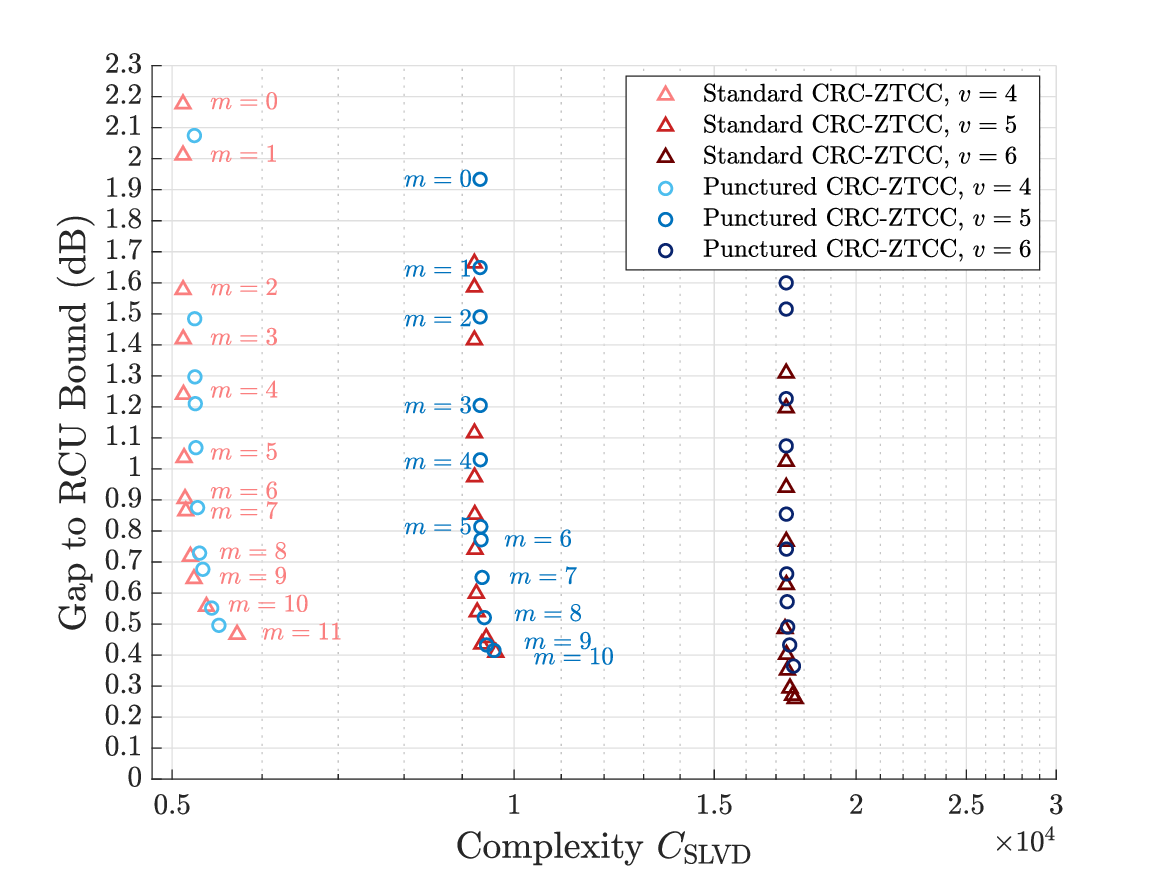}
\caption{The SNR gap to the RCU bound vs. the average complexity of SLVD of standard CRC-ZTCC codes in Table \ref{tab:ZT_codes_with_CRC_feedback} and punctured CRC-ZTCC codes in Table \ref{tab:ZT_codes_with_CRC_punc} for target FER of $10^{-4}$. Markers from top to bottom with the same color correspond to CRC polynomials with $m$ = $0$, $\dots$, $11$ for standard CRC-ZTCCs, and $m$ = $0$, $\dots$, $10$ for punctured CRC-ZTCCs. For punctured ZTCC with $v=4, m=0$, the gap to RCU bound is substantially high at $2.8262$ dB. }
\label{Fig:ztcc_gap2rcu}
\end{figure}

\subsection{Complexity Comparison}
\label{subsec: complexity comparison}

\begin{figure}
\centering
\includegraphics[width=20pc]{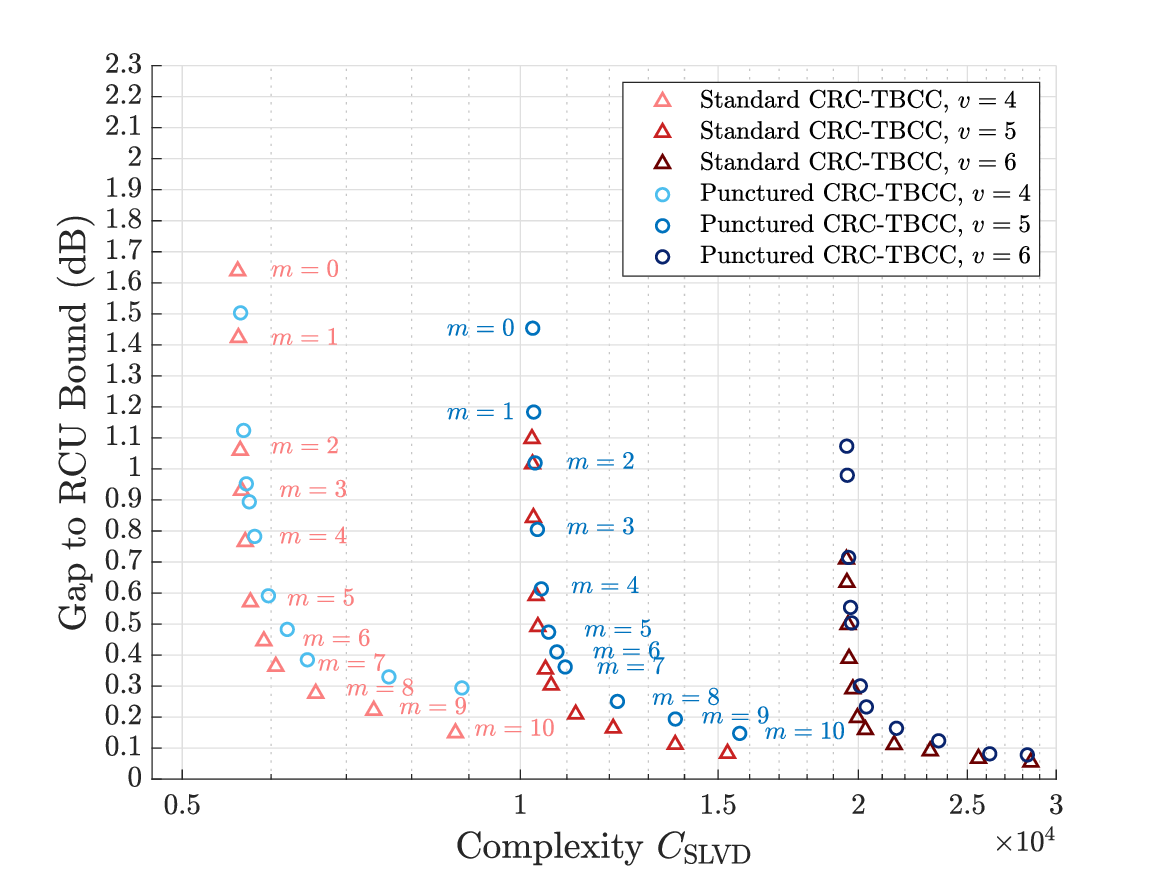}
\caption{The SNR gap to the RCU bound vs. the average complexity of SLVD of standard CRC-TBCC codes in Table \ref{tab:TB_codes_with_CRC_feedback} and punctured CRC-TBCC codes in Table \ref{tab:TB_codes_with_CRC_punc} for target FER of $10^{-4}$. Markers from top to bottom with the same color correspond to CRC polynomials with $m$ = $0$, $\dots$, $10$. For punctured TBCC with $v=4, m=0$, the gap to RCU bound is substantially high at $2.4257$ dB.}
\label{Fig:tbcc_gap2rcu}
\end{figure}

Fig. \ref{Fig:ztcc_gap2rcu} and \ref{Fig:tbcc_gap2rcu} display the trade-off between the SNR gap to the RCU bound and the average decoding complexity at the target FER of $10^{-4}$ for CRC-ZTCCs designed in Table \ref{tab:ZT_codes_with_CRC_feedback} and CRC-TBCCs designed in Table \ref{tab:TB_codes_with_CRC_feedback}. In addition, these figures directly compare the proposed dual trellis decoding scheme with the punctured scheme. The average decoding complexity of SLVD is evaluated according to the expressions in Sec. \ref{subsec: feedback complexity} and  \ref{subsec: punctured complexity}. We see that for a fixed $v$ (ZT or TB, standard or punctured), increasing the CRC degree $m$ significantly reduces the gap to the RCU bound, at the cost of a small increase in complexity. CRC-TBCCs generally have greater complexity than CRC-ZTCCs because the list decoder goes through many non-TB codewords.  The minimum gap of $0.25$ dB is achieved by the standard CRC-ZTCC with $v = 6$ and $m = 10$, and the minimum gap of $0.05$ dB is achieved by the CRC-TBCC with $v = 6$ and $m = 10$. For CRC-TBCCs, the gap to RCU bound continues to decrease when CRCs of higher degrees are applied, but the complexity grows substantially. For a more legible figure, we only show CRC polynomials of degrees up to $10$ in Fig. \ref{Fig:tbcc_gap2rcu}.

For the same CRC degree $m$, increasing the overall constraint length $v$ dramatically increases the complexity, while achieving a minimal reduction in the SNR gap to the RCU bound. On the other hand, the performance of CRC-ZTCC can be improved drastically by applying CRC polynomials of higher degrees. Both Fig. \ref{Fig:ztcc_gap2rcu} and \ref{Fig:tbcc_gap2rcu} demonstrate that for all three cases of constraint lengths $v$, one additional bit in the CRC benefits the decoding performance by moving closer to the RCU gap with a minimal cost in complexity. 

Additionally, for the same CRC degree $m$ and constraint length $v$, the standard high-rate codes generally perform better than the punctured codes while maintaining a similar decoding complexity. As the CRC degree increases, the performance and complexity of these two coding schemes draw nearer. Note that for CRC-ZTCCs, the rates for standard and punctured codes are different for $v=4$ and $v=5$, where the punctured codes have a higher rate due to fewer termination overhead bits.



\begin{figure}[t]
\centering
\includegraphics[width=20pc]{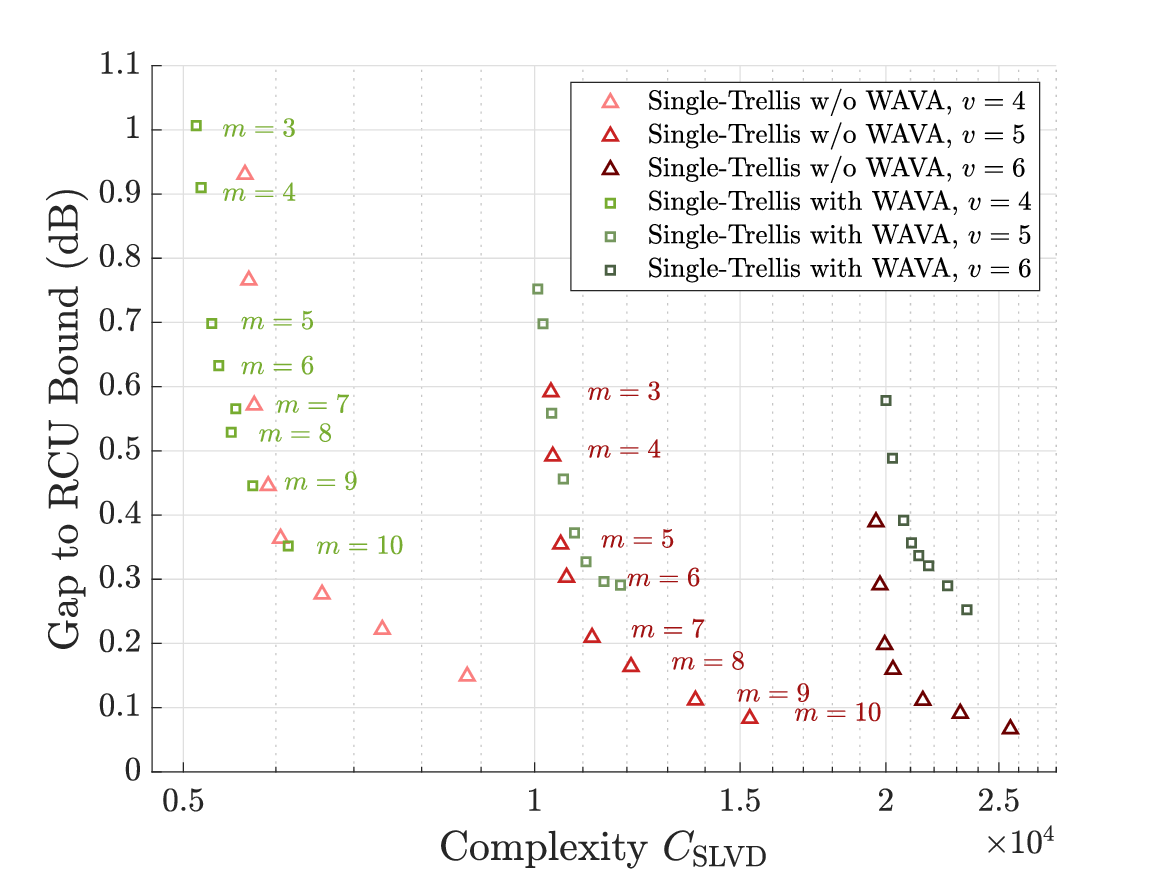}
\caption{The SNR gap to the RCU bound vs. the average complexity of SLVD of CRC-TBCC codes in Table \ref{tab:TB_codes_with_CRC_feedback} for target FER of $10^{-4}$. The results for both single-trellis decoding and WAVA decoding are demonstrated. Each color represents a specific CRC-aided CC shown in the table. Markers from top to bottom with the same color correspond to CRC polynomials with $m$ = $3$, $\dots$, $10$, where $m=0$ represents the convolutional codes without CRC.}
\label{Fig:WAVA_gap2rcu}
\end{figure}

Fig. \ref{Fig:WAVA_gap2rcu} shows the trade-off of complexity and performance for decoding CRC-TBCCs with a single trellis SLVD and a WAVA-based SLVD. The CRC-TBCCs used are of rate-$3/4$ for $m=3, \dots 10$ in \ref{tab:TB_codes_with_CRC_feedback}. The WAVA decoder has a  larger gap to RCU bound than the single-trellis decoder due to the extra ACS operations during the first traceback. However, the complexity of the WAVA decoder is smaller than that of the single-trellis decoder, and the difference increases as the CRC degree increases. For all constraint lengths $v$, the WAVA decoder at $m=10$ has a similar complexity as the single-trellis decoder at $m=6$.

\section{Results and Discussion}
\label{sec: results}
\begin{figure}[t]
\centering
\includegraphics[width=20pc]{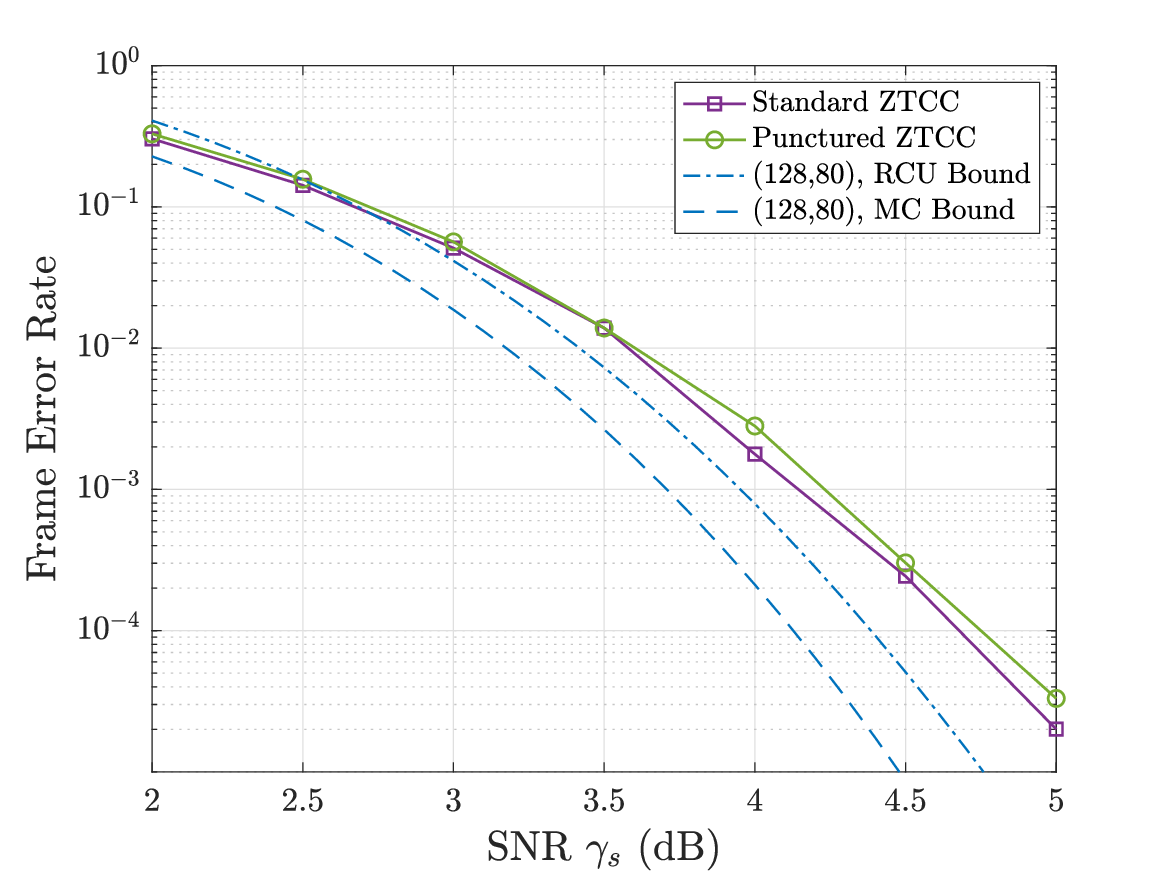}
\caption{FER vs. SNR for $v=6, m=10$ standard and punctured CRC-ZTCCs. The standard ZTCC is generated with the $(4,3,6)$ encoder $H=(107, 135, 133, 141)$ and the punctured ZTCC is generated by $G=(133,171)$ . The optimal CRC polynomials of degree  $10$ are 0x59F and 0x4E7, respectively. For the RCU and MC bounds, values in parenthesis denote blocklength $N$ and information length $K$, respectively. }
\label{Fig:ZT_FER_SNR_v6_N128}
\end{figure}

In this section, we will report and discuss the FER vs SNR performances of the CRC-CCs. Section \ref{subsec: ZTCC FER} compares the standard CRC-ZTCC with the punctured CRC-ZTCC, as well as shows the performance difference between using a single longer CRC and multiple shorter CRCs. Section \ref{subsec: TBCC FER} covers the performance of standard and punctured CRC-TBCCs.

\subsection{CRC-ZTCC Results}
\label{subsec: ZTCC FER}

Fig. \ref{Fig:ZT_FER_SNR_v6_N128} shows the performance comparison of standard and punctured $v=6$ CRC-ZTCCs with degree-$10$ CRC polynomials. For both CRC-ZTCCs, the blocklength is $128$ bits and the information length is $80$ bits, yielding a code rate of $0.625$. The standard CRC-ZTCC has slightly better FER performance than the punctured code. At the target FER of $10^{-4}$, the gap between the two schemes is around $0.08$ dB.

\begin{figure}[t]
\centering
\includegraphics[width=20pc]{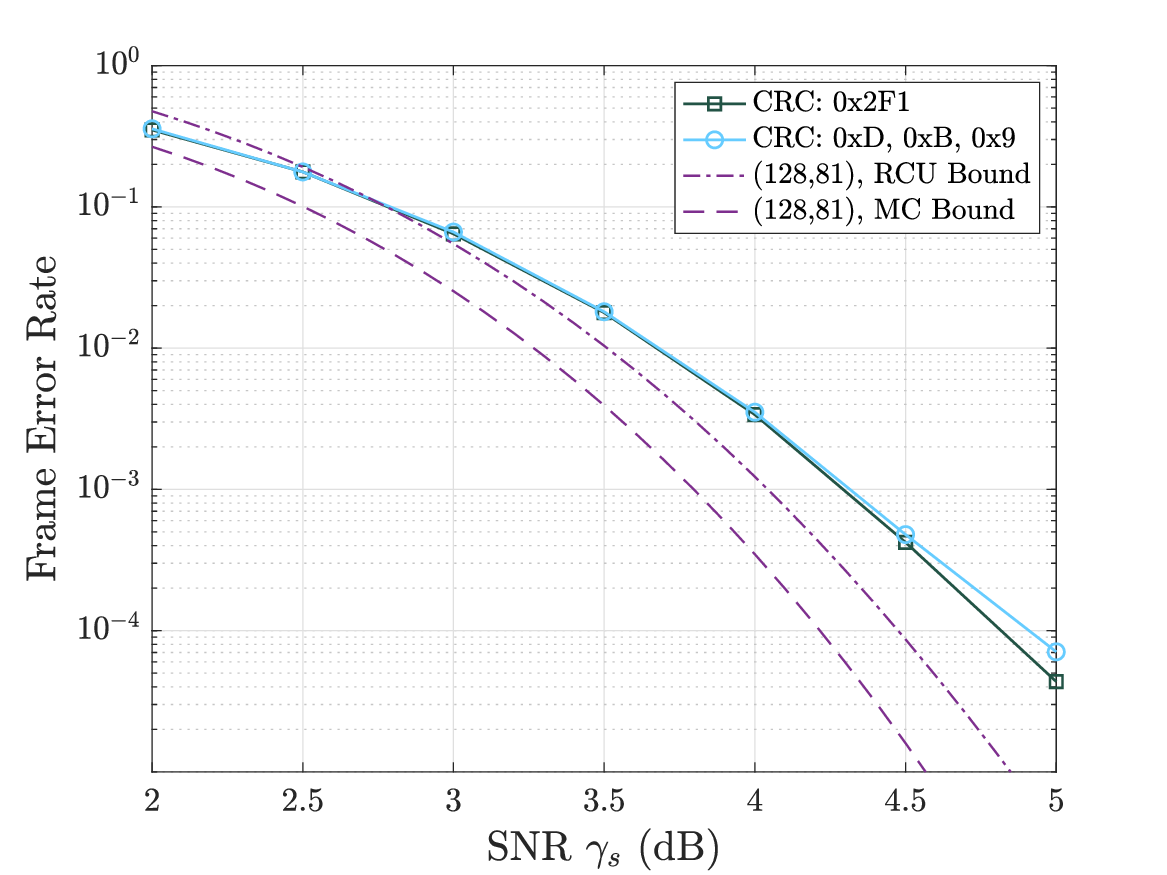}
\caption{FER vs. SNR for $v = 6$, $m=9$ CRC-ZTCCs designed under Karimzadeh \emph{et al.}'s scheme \cite{Karimzadeh2020} and our scheme. Both CRC-ZTCCs have information length $K=81$ and blocklength $N=128$. }
\label{Fig:ZT_FER_compare}
\end{figure}

In \cite{Karimzadeh2020}, Karimzadeh \emph{et al.} considered designing optimal CRC polynomials for each input rail of a multi-input CC. In their setup, an information sequence for an $(n, n-1, v)$ encoder needs to be split into $(n-1)$ subsequences before CRC encoding. In contrast, the entire information sequence in our framework is encoded with a single CRC polynomial. Then the resulting sequence is evenly divided into $(n-1)$ subsequences, one for each rail. To compare the performance between these two schemes, we design three degree-$3$ optimal CRC polynomials, one for each rail, for ZTCC with $H=(107,135,133,141)$. The three CRC polynomials jointly maximize the minimum distance of the CRC-ZTCC. For the single-CRC design, we use the single degree-$9$ optimal CRC polynomial for the same encoder from Table \ref{tab:ZT_codes_with_CRC_feedback}. Both CRC-ZTCCs have an information length $K=81$ and blocklength $N = 128$. Fig. \ref{Fig:ZT_FER_compare} shows the performance comparison between these two codes, showing that at high SNRs, a single degree-$9$ optimal CRC polynomial outperforms three degree-$3$ optimal CRC polynomials, one for each rail. This suggests that a single optimal CRC polynomial may suffice to provide superior protection for each input rail. The decoding complexity is similar regardless of the CRC scheme.

\subsection{CRC-TBCC Results}
\label{subsec: TBCC FER}
The performance of CRC-aided list decoding of ZTCCs relative to the RCU bound is constrained by the termination bits appended to the end of the original message, which are required to bring the trellis back to the all-zero state. TBCCs avoid this overhead by replacing the zero termination condition with the TB condition, which states that the final state of the trellis is the same as the initial state of the trellis \cite{Ma1986}.

\begin{figure}[t]
\centering
\includegraphics[width=20pc]{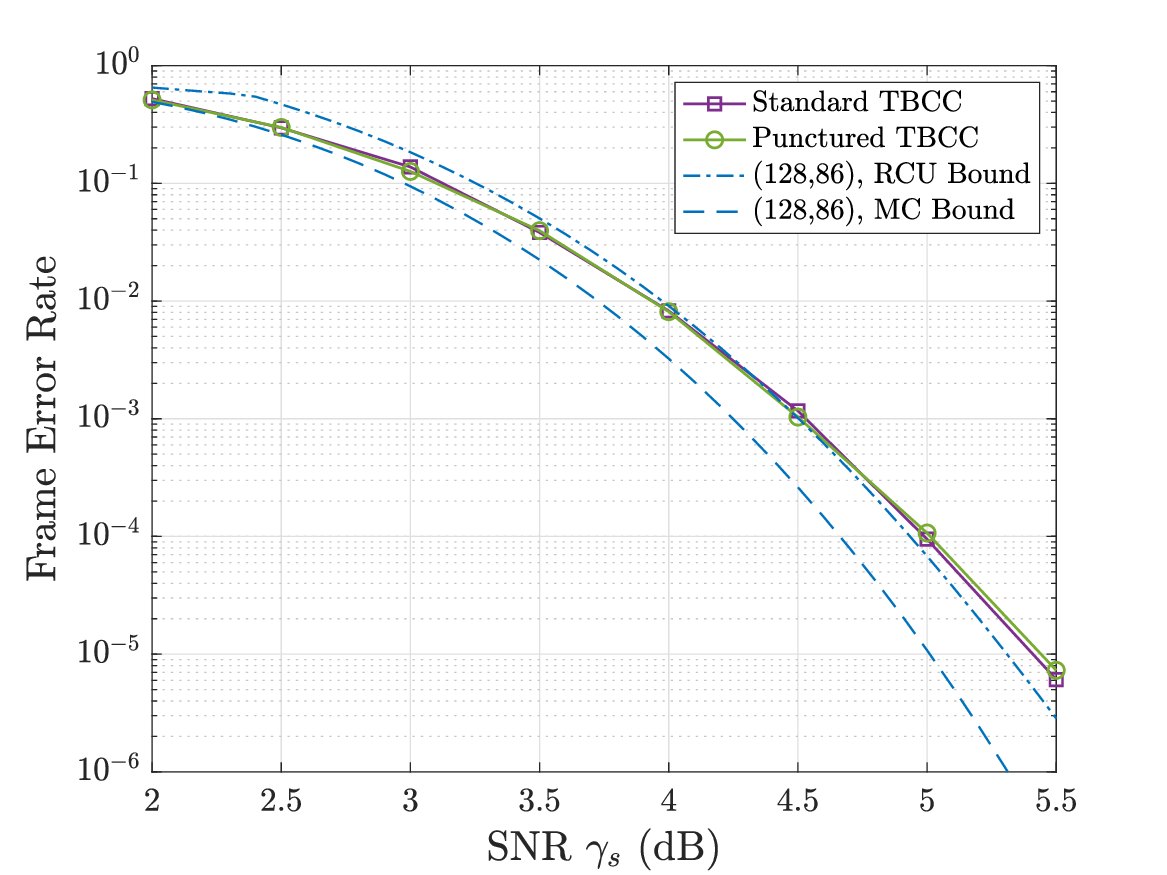}
\caption{FER vs. SNR for $v=6, m=10$ standard and punctured CRC-TBCCs. The standard TBCC is generated with the $(4,3,6)$ encoder $H=(107, 135, 133, 141)$ and the punctured TBCC is generated by $G=(133,171)$ . The optimal CRC polynomials of degree  $10$ are 0x529 and 0x40D, respectively. A single-trellis list decoder is used.}
\label{Fig:TB_FER_SNR_v6_N128}
\end{figure}

Fig. \ref{Fig:TB_FER_SNR_v6_N128} shows the FER vs. SNR for standard and punctured $v=6$ CRC-TBCCs with degree-$10$ CRC polynomials at a fixed blocklength of $128$ and information length of $86$. Both codes are able to closely approach the RCU bound. At the target FER of $10^{-4}$, the gap between the two schemes is within $0.05$ dB. Compared to the CRC-TBCCs with the same $v$ and $m$, the CRC-ZTCCs have a rate loss of around $0.04$ dB because of the termination overhead.

\section {Conclusion}
\label{sec: conclusion}
This paper shows that both standard and punctured high-rate CRC-aided CCs are able to approach the RCU bound for the BI-AWGN channel. The best CRC-TBCCs with the single-trellis ML decoder approach the RCU bound within $0.1$ dB for a target FER of $10^{-4}$ at a blocklength of $N=128$ bits. Concatenated with optimal CRC polynomials, the performance and complexity of the standard and punctured high-rate convolutional codes are similar.  In addition, adding one bit to the CRC can improve the FER  more than adding an additional memory element to the CC does for both standard and punctured CRC-CC schemes. 

For rate-$(n-1)/n$ TBCCs concatenated with optimal CRC polynomials, this paper considers three list decoding algorithms: a multi-trellis approach,  a single-trellis approach, and a modified single trellis approach with pre-processing using the Wrap Around Viterbi Algorithm (WAVA).   For the cases of standard codes, on which our simulations focused, all three algorithms use the dual trellis to reduce complexity.
The multi-trellis approach achieves the smallest expected list rank, but it suffers from a significantly larger overall complexity than the single-trellis approach. For the single trellis approach, we consider both an ML decoder and a non-ML decoder that uses WAVA pre-processing. WAVA pre-processing achieves a significantly smaller expected list size at the price of  a worse FER performance. 



\bibliographystyle{IEEEtran}
\bibliography{IEEEabrv,references}

\end{document}